\documentclass[aps,prd,twocolumn,showpacs,amsmath,amssymb,nofootinbib,eqsecnum, preprintnumbers]{revtex4-1}
\usepackage{graphicx}
\usepackage{epstopdf}
\usepackage{epsfig}

\def\appendix{{\newpage\section*{Appendix}}\let\appendix\section%
        {\setcounter{section}{0}
        \gdef\thesection{\Alph{section}}}\section}

\newcommand{\be}{\begin{equation}}
\newcommand{\ee}{\end{equation}}
\newcommand{\bear}{\begin{eqnarray}}
\newcommand{\eear}{\end{eqnarray}}
\newcommand{\ba}{\begin{array}}
\newcommand{\ea}{\end{array}}

\begin{document}

\title{Configurational entropy of tachyon kinks on unstable Dp-branes}
\author{Chong Oh Lee}
\email{cohlee@gmail.com}
\affiliation{Department of Physics, Kunsan
National University, Kunsan 573-701, Republic of Korea}

\begin{abstract}
We consider tachyon effective theory with Born-Infeld
electromagnetic fields and investigate the configurational entropy
of the various tachyon kink solutions. We find that the
configurational entropy starts at a minimum value and saturates to a
maximum value as the negative pressure of pure tachyonic field
increases. In particular, when an electric field is turned on and
its magnitude is larger than or equal to the critical value, we find
the configurational entropy has a global minimum, which is related
to the predominant tachyonic states.
\end{abstract}
\maketitle
\section{Introduction}
The investigation of  a D-brane with tachyon
condensation~\cite{Sen:1998sm} has opened up a new window to explore
off-shell structure of string theory. It was found that rolling
tachyon solutions through  boundary conformal field theory and
effective field theory describe the motion of the tachyon on
unstable D-branes~\cite{Sen:2002nu,Sen:2002in,Sen:2002an}. It was
extensively studied for the inhomogeneous tachyon
condensation~\cite{Cline:2002it,Felder:2002sv, Felder:2004xu,
Barnaby:2004nk}, the emission of closed string
radiation~\cite{Lambert:2003zr,Kluson:2003qk} and for instability of
codimension-one
D-branes~\cite{Lambert:2003zr,Kim:2003ina,Kim:2003ma, Sen:2003bc}.

It was found that the configurational entropy represents the
informational content in physical systems with localized energy
density configurations through measure of their ordering in field
configuration space~\cite{Gleiser:2011di,Gleiser:2012tu}. It was
extensively studied for AdS/QCD holographic models
~\cite{Bernardini:2016hvx,Bernardini:2016qit,Braga:2017fsb,Barbosa-Cendejas:2018mng,Karapetyan:2018yhm,Braga:2018fyc,Bernardini:2018uuy}
and  for instability of a variety of physical
systems~\cite{Gleiser:2012tu,Gleiser:2013mga,Gleiser:2014ipa,
Gleiser:2015rwa,Bernardini:2016hvx,Bernardini:2016qit,Casadio:2016aum,Braga:2016wzx,Lee:2017ero,Gleiser:2018kbq}.
In particular, the configurational entropy was investigated for
dynamical tachyonic AdS/QCD holographic model. The authors showed
that the corrections to dual mesonic states in the boundary QCD due
to tachyonic fields become more dominant their
states~\cite{Barbosa-Cendejas:2018mng}. Since decay of unstable
D-branes has rich tachyon kink solutions, it is intriguing to
calculate the configurational entropy of tachyon kink solutions.

The paper is organized as follows: in the next section we will
investigate configurational entropy in tachyon effective theory with
Born-Infeld electromagnetic fields. Firstly, we will calculate the
configurational entropy in the pure tachyon case. Next, we will turn
on electric/electromagnetic field. Then the configurational entropy
will be computed. We will also discuss their configurational entropy
by varying the electromagnetic field. In the last section we will
give our conclusion.

\section{Configurational entropy of tachyon kinks}
The Boltzmann-Gibbs entropy $S_{BG}$ is defined as
\bear\label{SBG}
S_{BG}=-k_B \sum p_i \ln p_i,
\eear
with $\sum p_i=1$.
In fact, it is given as the most general formula between the entropy and the set of probabilities of
their microscopic states in statistical thermodynamics.
Here, $k_B$ is the Boltzmann constant, and $p_i$ the probability of a microstate, respectively.
In particular, when each microstate has equal probability as the following
\bear\textstyle
p_i=\frac{1}{W},
\eear
with the number of microstates $W$, $S_{BG}$ (\ref{SBG}) reduces to
the configurational entropy $S_{\rm C}$ in the microcanonical ensemble
\bear\label{BS}
S_{\rm C}=k_B \ln W,
\eear
since $W$ can be treated as the number of possible configurations at a given energy.
For example, there are two different molecules with the total number of molecules $N_0$,
then the number of one type of molecule is $N_1$ and the number of another type of molecule
$N_2$.
One obtains the configurational entropy $S_{\rm C}$
\bear
S_{\rm C}=k_B \ln W=k_B \ln\left(\frac{N_0!}{N_1!\,N_2!}\right),
\eear
and after employing Sterling's approximation $\ln N!\approx N\ln N$, one has
\bear
S_{\rm C}=k_B(N_0\ln N_0-N_1\ln N_1-N_2\ln N_2).
\eear
As another example, there is the system with spatially localized energy in $d$-dimensional space.
When its energy density is given as a function of the position $\rho=\rho(x)$,
the energy density is written as
\bear\label{den}
\rho(k)=\left(\frac{1}{\sqrt{2\pi}}\right)^d \int \rho(x)e^{-ik\cdot x}d^dx
\eear
through the Fourier transform and the modal fraction reads
\bear\label{mf}
f(k)=\frac{|F(k)|^2}{\int|F(k)|^2d^dk},
\eear
which measures the relative weight of a given mode $k$.
One defines the configurational entropy $S_{\rm C}[f]$ as
\bear
S_{\rm C}[f]=-\sum_{l=1}^n f_l \ln(f_l).
\eear
and in the limit of $n\rightarrow\infty$, one has \cite{Gleiser:2013mga, Gleiser:2014ipa}
\bear\label{CS}
S_{\rm C}[f]=-\int_{-\infty}^{\infty} g(k) \ln [g(k)]d^dk,
\eear
where $g(k)=f(k)/f(k)_{\rm max}$ and the maximum modal fraction $f(k)_{\rm max}$.

One introduces a runaway tachyon potential\footnote{For dS tachyon thick braneworld model,
a more general tachyon potential $V(T)$ is given as~\cite{German:2012rv}
\bear\label{rp}
V(T)=-\Lambda_5~{\rm sech}(\sqrt{-2\kappa_5^2\Lambda_5/3}~T)
\sqrt{6~{\rm sech^2}(\sqrt{-2\kappa_5^2\Lambda_5/3}~T)-1}\nonumber
\eear
where $\Lambda_5$ is the five-dimensional bulk cosmological constant
and $\kappa_5$ is the five-dimensional gravitational coupling constant.}
\bear
V(T)=\frac{1}{\cosh(T/T_0)},
\eear
with $T_0=
\sqrt{2}$ for the non-BPS D-brane in the superstring and $T_0=2$ for the bosonic string.
The effective tachyon action for the unstable D$3$-brane system with the tension of the D$3$-brane ${\cal T}_3$
is given by \cite{Sen:1998sm, Sen:2002nu, Sen:2002in, Sen:2002an}
\bear\label{eta}\textstyle
S=-{\cal T}_3 \int d^{4}x V(T)\sqrt{-X},
\eear
with
\bear
X=\det(\eta_{\mu\nu}+\partial_{\mu} T \partial_{\nu}T+F_{\mu\nu})
\eear
which leads to equations of motion for the gauge field $A_\mu$ and for the tachyon $T$
\bear\label{gaes}
\partial_{\mu}\left(\frac{V(T)}{\sqrt{-X}}C_{\rm A}^{\mu\nu}\right)=0,
\eear
\bear
\partial_{\mu}\left(\frac{V(T)}{\sqrt{-X}}C_{\rm S}^{\mu\nu}\partial_{\nu}T\right)+\sqrt{-X}\frac{dV(T)}{dT}=0.
\eear
Here $C_{\rm A}^{\mu\nu}$, and $C_{\rm S}^{\mu\nu}$ are asymmetric and symmetric part of the cofactor,
\bear
C_{\rm A}^{\mu\nu}=\bar{\eta}(\bar{F}^{\mu\nu}+\bar{\eta}^{\mu\alpha}
\bar{\eta}^{\beta\gamma}\bar{F}_{\alpha\beta}^*\bar{F}_{\gamma\delta}^*\bar{F}^{\delta\nu}),
\eear
\bear
C_{\rm S}^{\mu\nu}=\bar{\eta}(\bar{\eta}^{\mu\nu}+\bar{\eta}^{\mu\alpha}
\bar{\eta}^{\beta\gamma}\bar{\eta}^{\delta\nu}\bar{F}_{\alpha\beta}^*\bar{F}_{\gamma\delta}^*)
\eear
with determinant of barred metric $\bar{\eta}$
\bear
\bar{\eta}=-(1+\partial_\mu T \partial^\mu T),
\eear
and inverse metric $\bar{\eta}^{\mu\nu}$
\bear
\bar{\eta}^{\mu\nu}=\eta^{\mu\nu}-\frac{\partial^\mu T \partial^\nu T}{1+\partial_\alpha T \partial^\alpha T}.
\eear
Here, contravariant barred field strength tensor $\bar{F}^{\mu\nu}$ denotes
\bear
\bar{F}^{\mu\nu}=\bar{\eta}^{\mu\alpha}\bar{\eta}^{\nu\beta}\bar{F}_{\alpha\beta},
\eear
and barred field strength tensor  $\bar{F}_{\mu\nu}$
\bear
 \bar{F}_{\mu\nu} =F_{\mu\nu},
\eear
and its dual field strength $\bar{F}_{\mu\nu}^*$
\bear
\bar{F}_{\mu\nu}^*=\frac{1}{2}{\bar\epsilon}_{\mu\nu\alpha\beta}\bar{\eta}^{\alpha\gamma}\bar{\eta}^{\beta\delta}\bar{F}_{\gamma\delta}.
\eear
Energy-momentum tensor $T_{\mu\nu}$ is obtained as~\cite{Kim:2003ina, Kim:2003ma}
\bear
T^{\mu\nu}={\cal T}_3\frac{V(T)}{\sqrt{-X}}C_{\rm S}^{\mu\nu}.
\eear
One considers the tachyon field $T(x)$ as the function of the spatial coordinate $x$, electric field ${\bold E}(x)$
and the magnetic field  ${\bold B}(x)$.
Then through solving Eq. (\ref{gaes}), one obtains that
\bear
\tau\equiv {\cal T}_3 \frac{V(T)}{\sqrt{-X}},
\eear
with $\tau=constant$, which leads to a single first-order equation
\bear\label{feq}
{\cal E}=\frac{1}{2}T^{\prime2}+U(T),
\eear
with
\bear\label{me}
{\cal E}=-\kappa/(2\lambda)
\eear
and
\bear\label{po}
U(T)=-({\cal T}_3V(T))^2/(2\lambda \tau^2).
\eear
Here $\kappa$ and $\lambda$ are
\bear
\kappa&=&1-{\bold E}^2+{\bold B}^2-({\bold E}\cdot{\bold B})^2,\\
\label{lam}\lambda&=&1+B_1^2-E_2^2.
\eear

\subsection{Tachyon kink case}
When the tachyon field and the abelian gauge field are given as the function of the position $T(x)$ and $F_{\mu\nu}(x)$,
profiles of tachyon field $T(x)$ on general unstable D$p$-brane are the same to that on unstable D$2$-brane~\cite{Kim:2003ma}.
For simplicity, we consider from now on
tachyon effective theory with Born-Infeld electromagnetic fields in three-dimensional
spacetime and investigate the configurational entropy of the various tachyon kink solutions.

In the case of pure tachyon model, the Born-Infeld type effective
action (\ref{eta}) reduces to the following action\footnote{One
considers generalization of the action to include the bulk action
with the five-dimensional bulk cosmological constant
\bear
S&=&\int d^5x \sqrt{-\det(g_{\mu\nu})}(R/2\kappa_5^2-\Lambda_5)\nonumber\\
&& -\int d^5x V(t)
\sqrt{-\det(g_{\mu\nu}+\partial_{\mu}T\partial_{\nu}T)},\nonumber
\eear
which is adapted to a warped five-dimensional line element
with an induced three-dimensional brane in a spatially flat
cosmological background
\bear
ds^2=e^{2f(\xi)}[-dt^2+e^{2Ht}(dx^2+dy^2+dz^2)+d\xi^2],\nonumber
\eear
with the warp factor $f(\xi)$ and the scale factor of the brane
$e^{Ht}$, where $H$ is integration constant, and one can get the
following field equation~\cite{German:2012rv}:
\bear
f^{'2}+\frac{\kappa_5^2\Lambda_5}{6}e^{2f}-H^2=-\kappa_5^2\frac{e^{2f}V(T)}{6\sqrt{1+e^{-2f}T^{'2}}}.\nonumber
\eear
The five-dimensional gravitational coupling constant $\kappa_5$ is set to $\sqrt{6}$ and the above field equation
in the absence of the negative bulk cosmological constant $\Lambda_5$ and the warp factor $f(\xi)$
reduces to the negative pressure (\ref{nep}) $p_1=-\frac{V(T)}{\sqrt{1+T^{\prime2}}}$
with taking the tension of the D2-brane ${\cal T}_2=1$.
Then the more general potential discussed in footnote 1, reduces exactly to the runaway tachyon potential
(\ref{rp}) $V(T)=\frac{1}{\cosh(T/T_0)}$.
}:
\bear\label{ptact}\textstyle
S=-{\cal T}_2 \int d^{3}x V(T)\sqrt{-\det(\eta_{\mu\nu}+\partial_{\mu} T \partial_{\nu}T)},
\eear
which leads to the energy density $\rho$~\cite{Kim:2003ina, Kim:2003ma}
\bear
\rho\equiv T_{00}=\frac{-{\cal T}_2^2/p_1}{1+\left[(\frac{{\cal T}_2}{p_1})^2-1\right]\sin^2(x/T_0)},
\eear
and the negative pressure $p_1$
\bear\label{nep}
p_1\equiv T_{11}=-{\cal T}_2\frac{V(T)}{\sqrt{1+T^{\prime2}}}<0.
\eear
The mechanical energy ${\cal E}$ (\ref{me}) is given as ${\cal E}=-1/2$ and the
potential $U(T)$ (\ref{po}) is $U(T)=-({\cal T}_2 V(T))^2/2$. Then,
solutions are classified by $-p_1/{\cal T}_2$ and there are four
possible cases: (i) When $-p_1/{\cal T}_2>1$, there is no motion.
(ii) When $-p_1/{\cal T}_2=1$, there is hypothetical motion but the
motion eternally stops at $U(0)$. (iii) When $-p_1/{\cal T}_2<1$,
there is oscillatory motion. (iv) When $-p_1/{\cal T}_2\rightarrow
0^+$, there is oscillatory motion with infinity period. Thus, we
now calculate  the configurational entropy of pure tachyonic field
under $0<-p_1/{\cal T}_2<1$.

The potential $U(T)$ (\ref{po}) is explicitly obtained as
\bear
U(T)=-\frac{{\cal T}_2^2}{2p_1^2}{\rm sech}^2\left(\frac{T}{T_0}\right).
\eear
After taking ${\cal T}_2=1$, shapes of potential $U(T)$
for various values of the pressure $p_1$ are depicted in Fig. 1.
The smaller the pressure $-p_1$ becomes , the more convex function in potential $U(T)$.
Then, profiles of tachyon filed  $T(x)$ for various the pressure $-p_1$ and
profiles of energy density are depicted in Fig. 2 and Fig. 3.

\begin{figure}[!htbp]
\begin{center}
{\includegraphics[width=8cm]{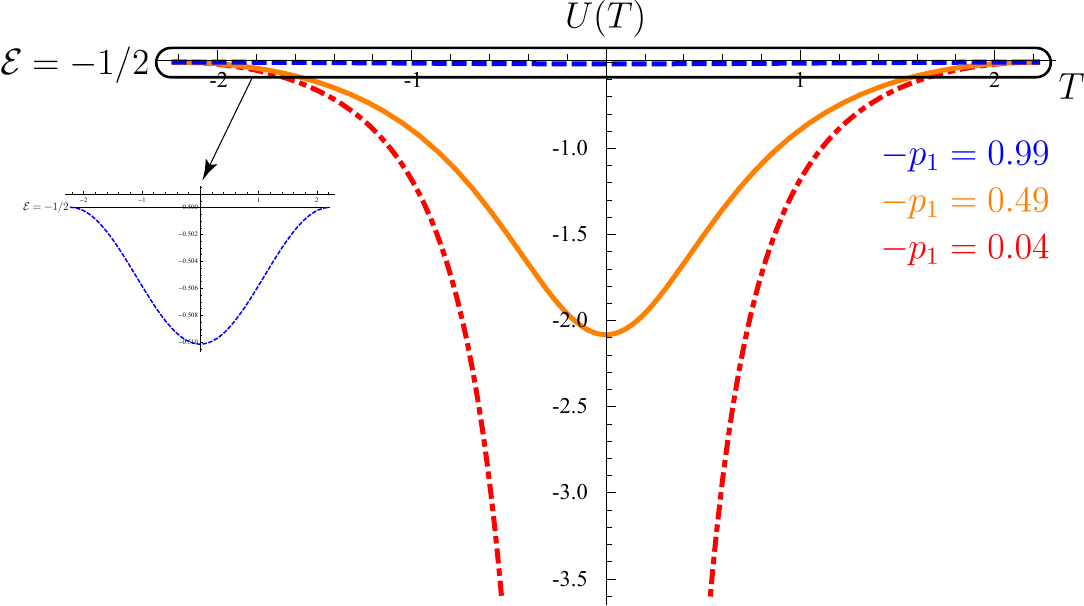}}
\end{center}
\vspace{-0.6cm}
\caption{{\footnotesize Plot of potential $U(T)$ as
the function of the tachyon field $T$ (red dotted-dashed curve for
the pressure $-p_1 =0.04$, orange solid curve for the pressure $-p_1
=0.49$, blue dashed curve for the pressure $-p_1 =0.99$,
respectively).}}
\label{figI}
\end{figure}

\begin{figure}[!htbp]
\begin{center}
{\includegraphics[width=8cm]{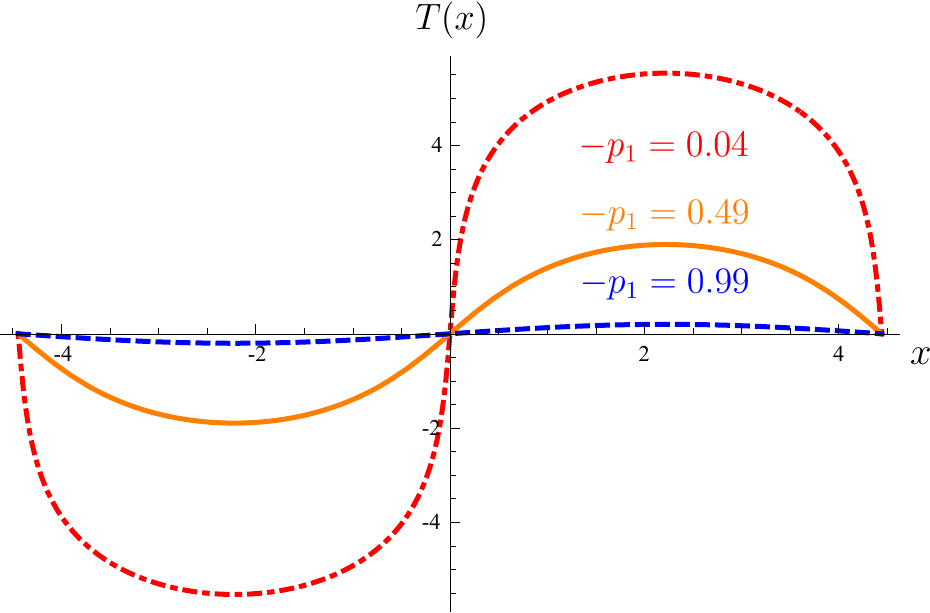}}
\end{center}
\vspace{-0.6cm}
\caption{{\footnotesize Plot of the tachyon field $T(x)$ as
the function of the position $x$ (red dotted-dashed curve for
the pressure $-p_1 =0.04$, orange solid curve for the pressure $-p_1
=0.49$, blue dashed curve for the pressure $-p_1 =0.99$,
respectively).}}
\label{figII}
\end{figure}

\begin{figure}[!htbp]
\begin{center}
{\includegraphics[width=8cm]{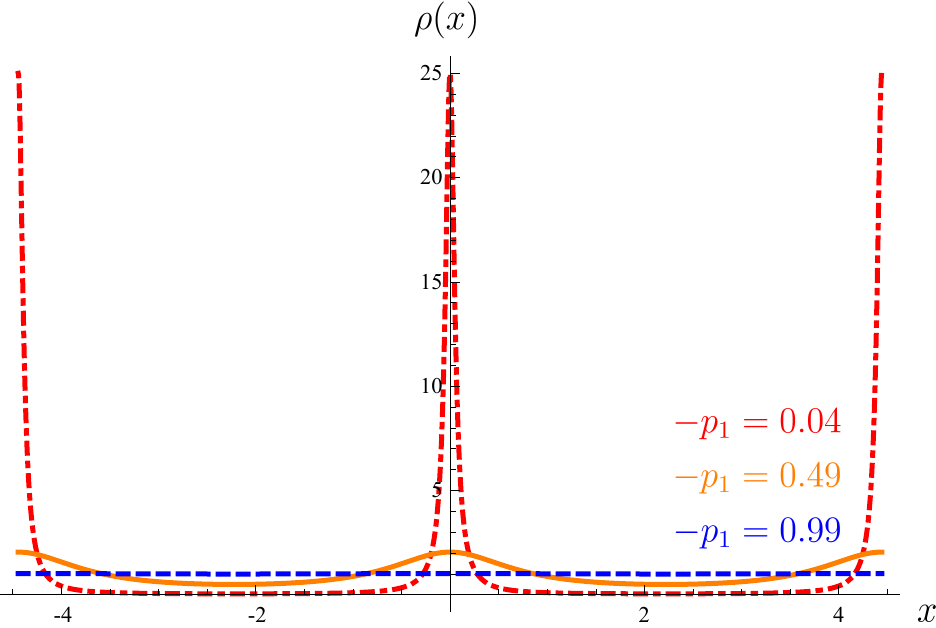}}
\end{center}
\vspace{-0.6cm}
\caption{{\footnotesize Plot of energy density $\rho(x)$ as
the function of the position $x$ (red dotted-dashed curve for
the pressure $-p_1 =0.04$, orange solid curve for the pressure $-p_1
=0.49$, blue dashed curve for the pressure $-p_1 =0.99$,
respectively).}}
\label{figIII}
\end{figure}

The configurational entropy of tachyon filed $S_{\rm C,T}$ is numerically calculated
by Eqs. (\ref{den}), (\ref{mf}), and (\ref{CS}), and
is depicted in Fig. 4. As the negative pressure $-p_1$ grows up,
$S_{\rm C,T}$ starts at the minimum value ($S_{\rm C,T}=0.5441$)
and saturates to the maximum value ($S_{\rm C,T}=1.7373$).

The momentum space plane waves with equally distributed modal
fractions are sharply localized in position space while the momentum
space singular modes broadly spreads out. Position space localized
distributions have the maximum configurational entropy due to large
amount of momentum modes while position space widespread
distributions have the minimum configurational entropy due to small
amount of momentum modes. Furthermore, the larger configurational
entropy of physical system becomes, the larger its amount of energy
to be generated. Thus in position space, the energy density with
more sharply localized shape needs a larger amount of energy to be
generated. On the contrary, the more energy density $\rho(x)$ in the
presence of pure tachyonic fields  sharply localized, the smaller
the pressure $-p_1$ becomes while the more energy density $\rho(x)$
broadly spreads out, the bigger the pressure $-p_1$ becomes as shown
in Fig. 3. As shown in Fig. 4, however, the above behaviour of
configurational entropy by increasing the pressure $-p_1$ seems to
be natural physically since the tachyon energy density with more
widespread shape needs a larger amount of energy to be generated. In
other words, this result is consistent with the fact that the
configurational entropy grows up  as the energy increases.

\begin{figure}[!htbp]
\begin{center}
{\includegraphics[width=8cm]{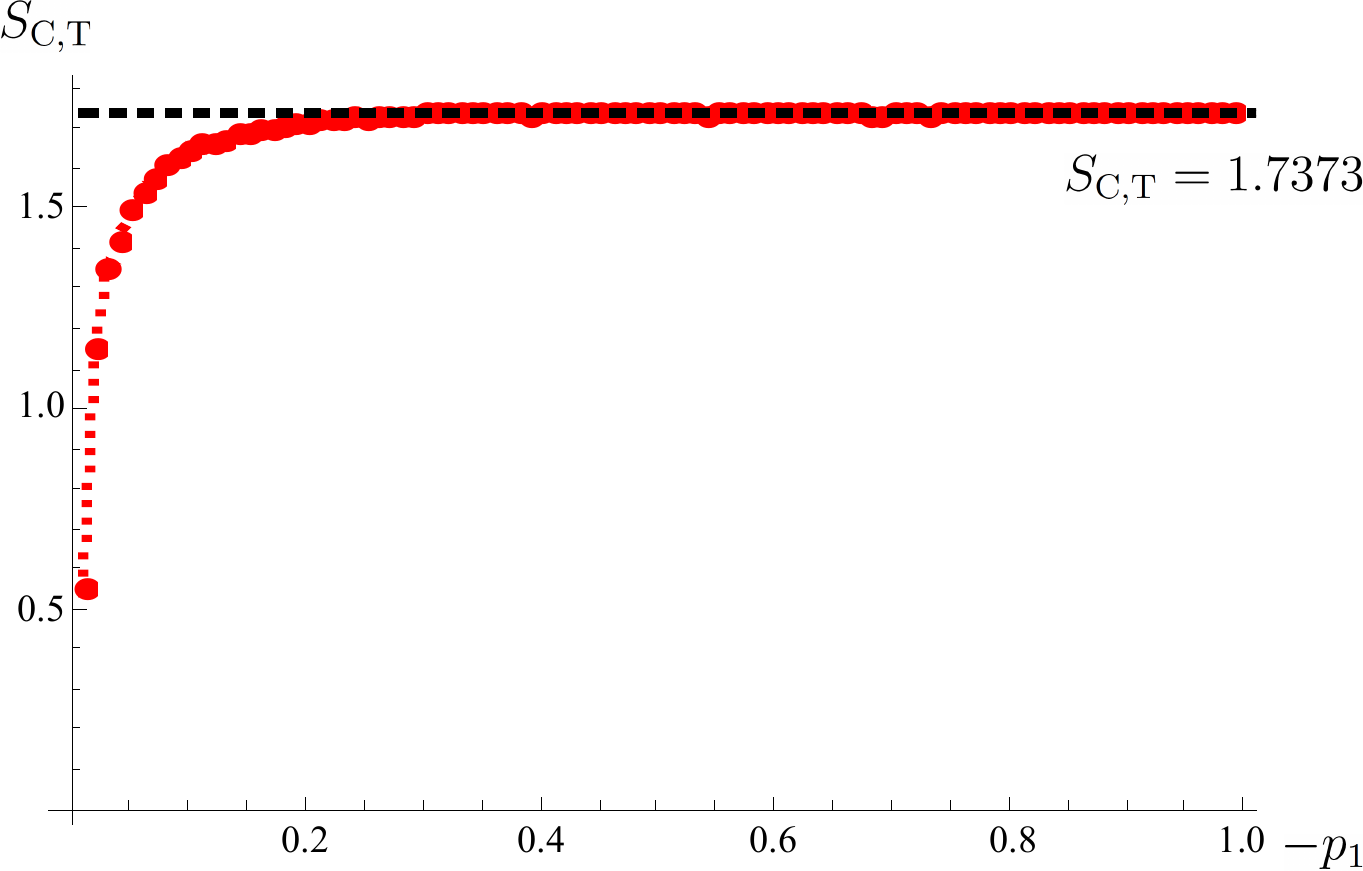}}
\end{center}
\vspace{-0.6cm}
\caption{{\footnotesize Plot of tachyon configurational entropy $S_{\rm C,T}$
as the function of the pressure $-p_1$.}}
\label{figIV}
\end{figure}

\subsection{Kink with electric field case}
As discussed in the previous section we will apply a similar analysis
to the kink with electric field case.

The energy density $\rho$ is given as~\cite{Kim:2003ina, Kim:2003ma}
\bear
\rho&\equiv& T_{00}=\Pi E\nonumber\\
&&+\frac{{\cal T}_2^2E/\Pi}
{1+\left(\frac{{\cal T}_2^2E^2}{\Pi^2(1-E^2)}-1\right)\sin^2(x\sqrt{1-E^2}/T_0)},\nonumber\\
\eear
and conjugate momenta of the gauge field $\Pi$
\bear
\Pi={\cal T}_2\frac{V(T)}{\sqrt{-X}}E,
\eear
where $E$ denotes the electric field.
The mechanical energy (\ref{me}) is given as
\bear
{\cal E}_{\rm E}=-\frac{1}{2}(1-E^2),
\eear
and the potential (\ref{po}) is
\bear
U_{\rm E}(T)&=&-\frac{{\cal T}_2E^2}{2\Pi^2}V(T)^2\nonumber\\
&=&-\frac{{\cal T}_2E^2}{2\Pi^2}\frac{1}{\cosh^2(T/T_0)}.
\eear
Then, solutions are classified as ${\cal E}$ or $E$. (i) When ${\cal
E}_{\rm E} <U_{\rm E}(0)~(E^2>1/[1+({\cal T}_2/\Pi)^2])$, there is
no motion. (ii) When ${\cal E}_{\rm E} =U_{\rm E}(0)~(E^2=1/[1+({\cal
T}_2/\Pi)^2])$, there is hypothetical motion but the motion
eternally stops at U(0). (iii) $U_{\rm E}(0)<{\cal E}_{\rm
E}<0~(1/[1+({\cal T}_2/\Pi)^2]<E^2<1)$, there is oscillatory motion.
Thus, we will calculate the configurational entropy $S_{\rm C,E}$ in
the range $0<E<1$.

After employing ${\cal T}_2=1$, shapes of potential $U_{\rm E}(T)$
for various values of electric field $E$ are depicted in Fig. 5.
The smaller electric field $E$ becomes, the more convex function in potential $U_{\rm E}(T)$.
Then, profiles of tachyon filed  $T(x)$ for various electric field $E$ and
profiles of energy density are depicted in Fig. 6 and Fig. 7.

\begin{figure}[!htbp]
\begin{center}
{\includegraphics[width=8cm]{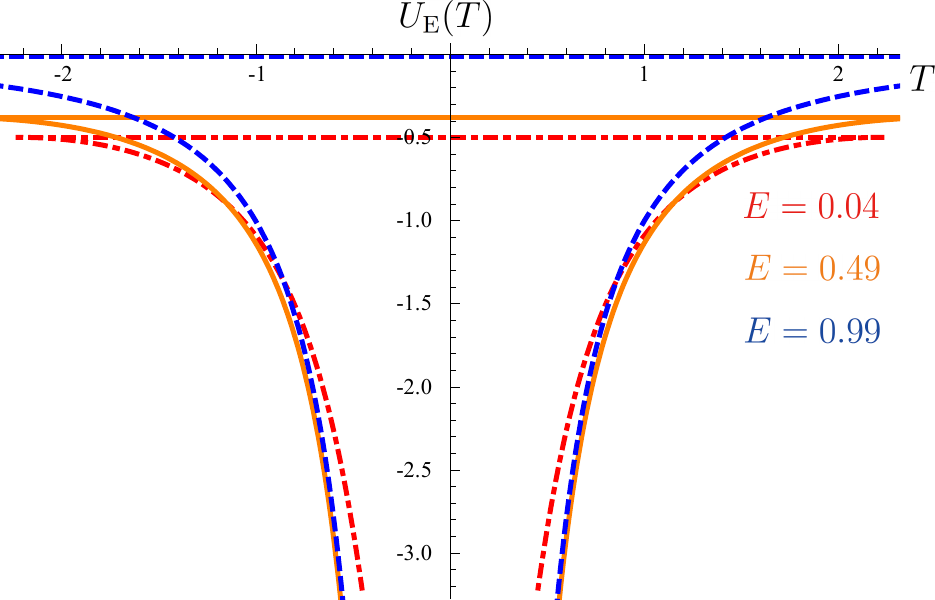}}
\end{center}
\vspace{-0.6cm}
\caption{{\footnotesize Plot of potential $U_{\rm E}(T)$ as
the function of the tachyon field $T$ (red dotted-dashed curve for
electric field $E=0.04$, orange solid curve for electric field $E
=0.49$, blue dashed curve for electric field $E=0.99$,
respectively).}}
\label{figV}
\end{figure}

\begin{figure}[!htbp]
\begin{center}
{\includegraphics[width=8cm]{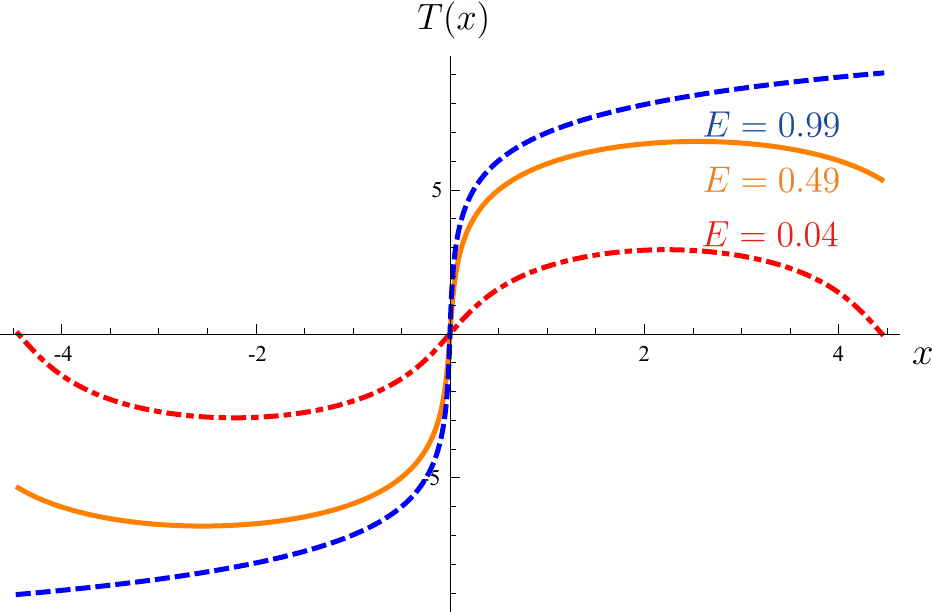}}
\end{center}
\vspace{-0.6cm}
\caption{{\footnotesize Plot of the tachyon field $T(x)$ as
the function of the position $x$ (red dotted-dashed curve for
electric field $E =0.04$, orange solid curve for electric field $E
=0.49$, blue dashed curve for electric field $E=0.99$,
respectively).}}
\label{figVI}
\end{figure}

\begin{figure}[!htbp]
\begin{center}
{\includegraphics[width=8cm]{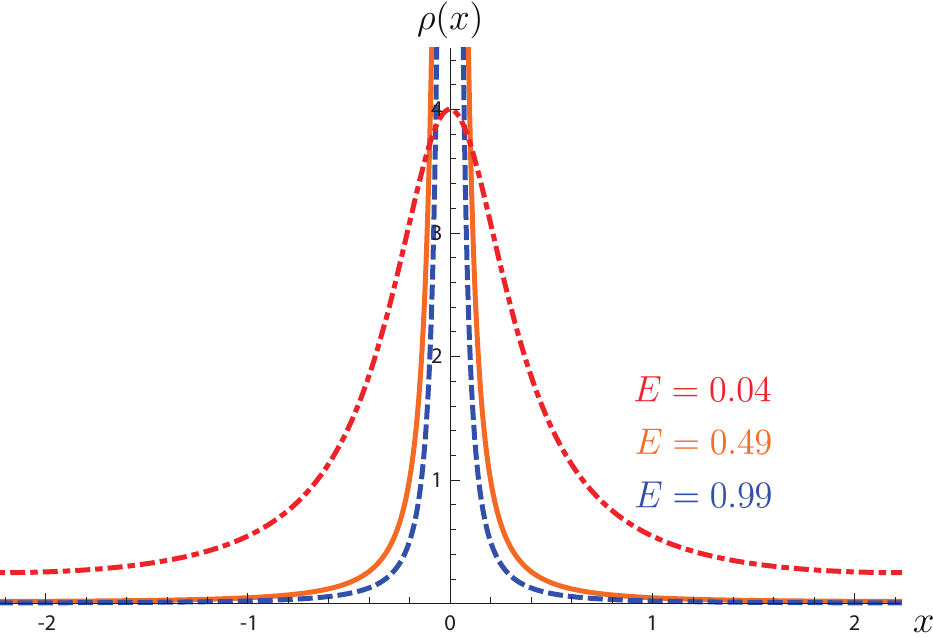}}
\end{center}
\caption{{\footnotesize Plot of energy density $\rho(x)$ as
the function of the position $x$ (red dotted-dashed curve for
electric field $E =0.04$, orange solid curve for electric field $E
=0.49$, blue dashed curve for electric field $E=0.99$,
respectively).}}
\label{figVII}
\end{figure}

\begin{figure}[!htbp]
\begin{center}
{\includegraphics[width=8cm]{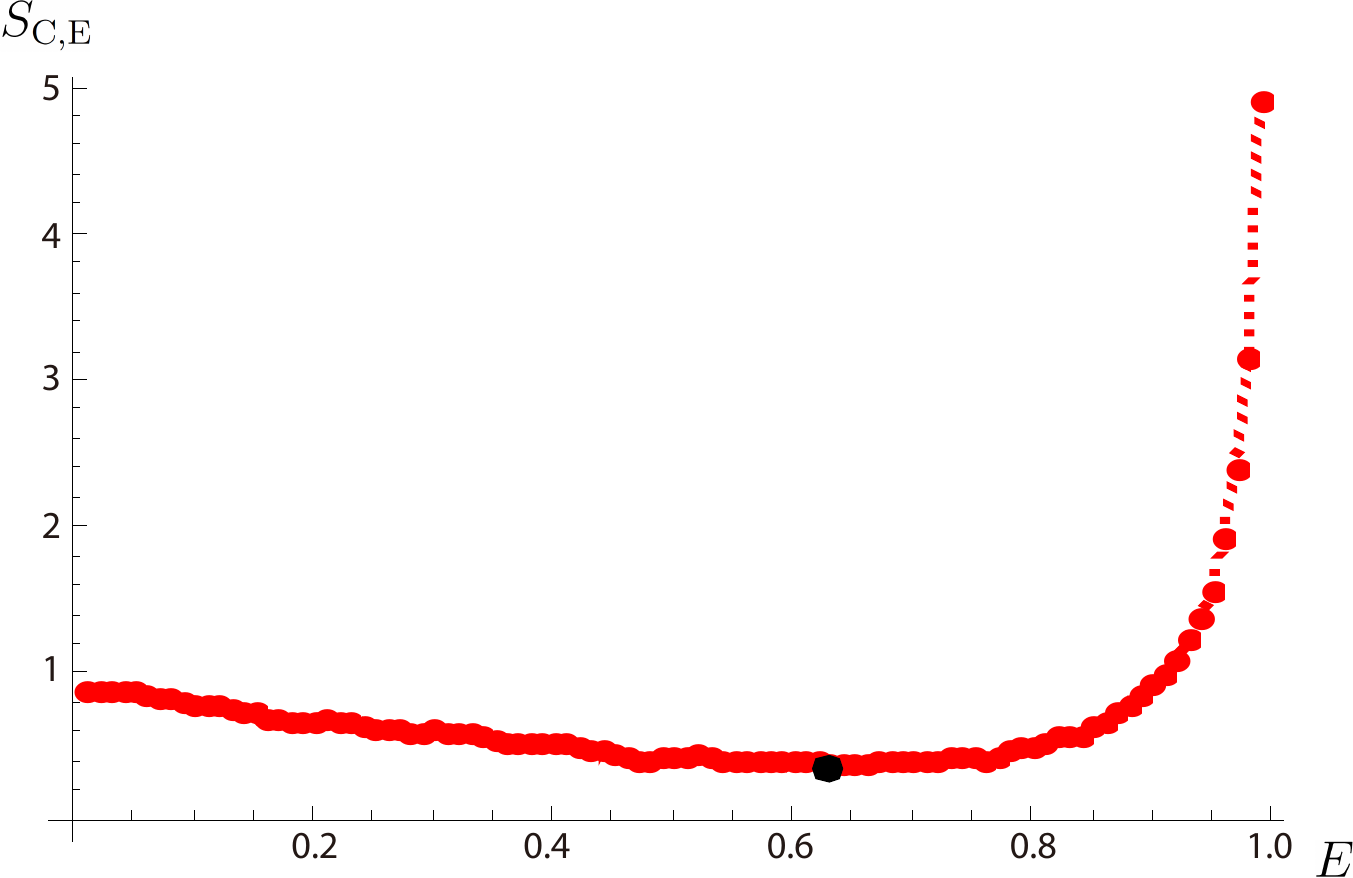}}
\end{center}
\vspace{-0.6cm}
\caption{{\footnotesize Plot of tachyon configurational entropy
with electric field $S_{\rm C,E}$
as the function of electric field $E$.}}
\label{figVIII}
\end{figure}

After numerically evaluating Eqs. (\ref{den}), (\ref{mf}), and (\ref{CS}),
the configurational entropy $S_{\rm C,E}$ is depicted in Fig. 8.
Especially, as electric field $E$ grows up, the configurational entropy
reaches the minimum value ($S_{\rm C,E}=0.3485$) at a critical point ($E=0.63$).
In fact, the smaller configurational entropy of physical system becomes,
the more dominant such physical system states
since the larger its configurational entropy becomes,
the larger its amount of energy to be generated.
Thus, it is expected that the predominant tachyonic states
happens at the minimum configurational entropy.

\subsection{Kink with electromagnetic field case}
In the case of the kink with electromagnetic field,
the energy density $\rho$ is given as~\cite{Kim:2003ina, Kim:2003ma}
\bear\label{Emrho}
\rho&\equiv& T_{00}={\cal T}_2\frac{V(T)}{\sqrt{-X}}(1+T^{\prime2}+B^2)\nonumber\\
&=&\frac{\Pi_1(E_1^2-B^2E_2^2)}{E_1(1-E_2^2)}+\frac{{\cal T}_2^2E_1}{\Pi_1(1-E_2^2)}V(T)^2.
\eear
The mechanical energy (\ref{me}) is given as
\bear
{\cal E}_{\rm EM}=-\frac{1-{\bold E}^2+B^2}{2(1-E_2^2)},
\eear
where electric field ${\bold E}$ is given as
\bear
{\bold E}=E_1(x) \hat{i}+E_2(x) \hat{j}.
\eear
Three-dimensional analogue of Faraday's law is
\bear
\frac{\partial B}{\partial t}=-\epsilon_{0ij}\partial_i E_,
\eear
which implies magnetic field $B$
\bear
B=B(x)
\eear
where $E_i=F_{i0}$, and $B=\epsilon_{0ij}F_{ij}/2$.

The potential (\ref{po}) is
\bear
U_{\rm EM}(T)&=&-\frac{{\cal T}_2E_1^2}{2\Pi_1^2(1-E_2^2)}V(T)^2\nonumber\\
&=&-\frac{{\cal T}_2E_1^2}{2\Pi_1^2(1-E_2^2)}\frac{1}{\cosh^2(T/T_0)}.
\eear
When (\ref{lam}) $\lambda>0$ $(E_2<1)$, tachyon configurations are
exactly same to that in kink with electric field case. However, when
$\lambda<0$ $(E_2>1)$, the potential $U_{\rm EM}$ is flipped, the
property of solutions changes. Thus, we will treat the case in which
$\lambda$ is negative.

\begin{figure}[!htbp]
\begin{center}
{\includegraphics[width=8cm]{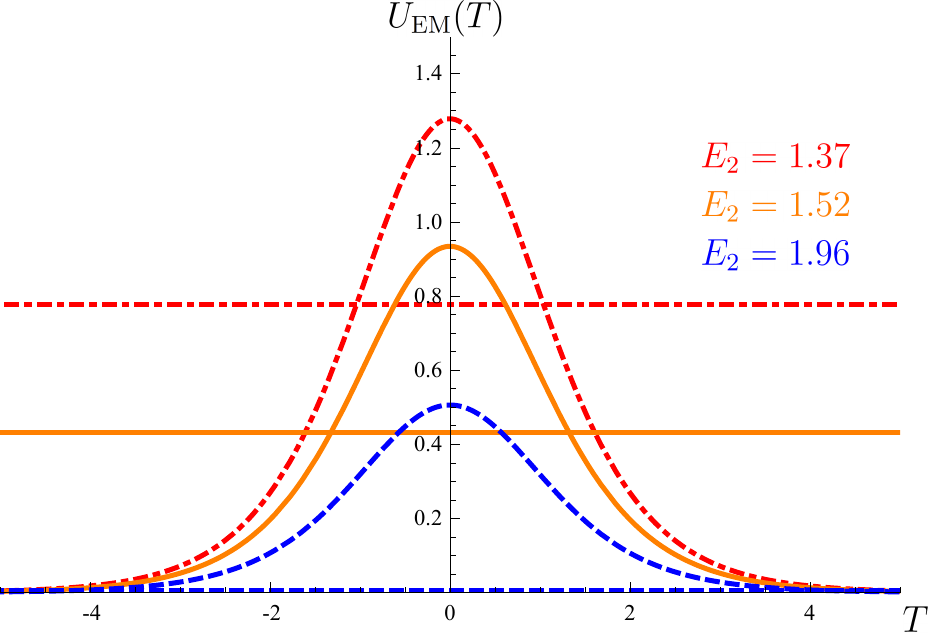}}
\end{center}
\vspace{-0.6cm}
\caption{{\footnotesize Plot of potential $U(T)$ as
the function of the tachyon field $T$ (red dotted-dashed curve for
electric field $E_2=1.37$, orange solid curve for electric field $E_2
=1.52$, blue dashed curve for electric field $E_2=1.96$,
respectively).}}
\label{figIX}
\end{figure}

\begin{figure}[!htbp]
\begin{center}
{\includegraphics[width=8cm]{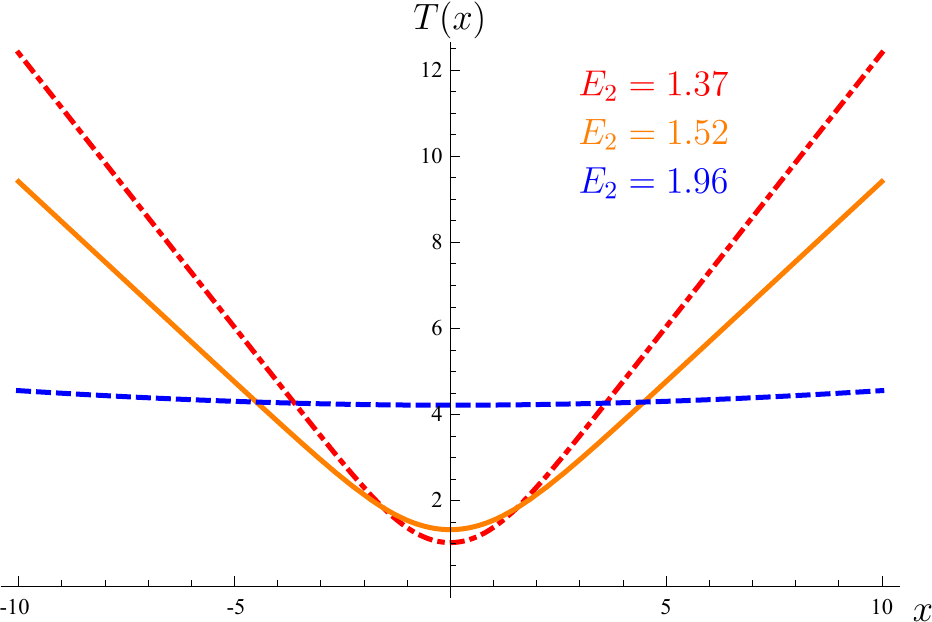}}
\end{center}
\vspace{-0.6cm}
\caption{{\footnotesize Plot of the tachyon field $T(x)$ as
the function of the position $x$(red dotted-dashed curve for
electric field $E_2=1.37$, orange solid curve for electric field $E_2
=1.52$, blue dashed curve for electric field $E_2=1.96$,
respectively).}}
\label{figXI}
\end{figure}

(i) When $0<{\cal E}_{\rm EM}<U_{\rm EM}(0)$ $(0<1-{\bold E}^2+B^2<({\cal T}_2E_1/\Pi_1)^2)$,
there is a turning point at the origin ($x=0$) as shown in Fig 10.
Then, the charge density is explicitly written as
\bear
\rho&=&\frac{\Pi_1(B^2E_2^2-E_1^2)}{E_1(E_2^2-1)}\nonumber\\
&&-\frac{{\cal T}_2^2E_1}{\Pi_1(E_2^2-1)}\frac{1}{1+(\chi^2-1)\cosh^2(x/\sigma)},
\eear
with $\chi$
\bear
\chi=\frac{{\cal T}_2^2E_1^2}{\Pi_1^2|1-{\bold E}^2+B^2|},
\eear
and $\sigma$
\bear
\sigma=T_0\sqrt{\left|\frac{1-E_2^2}{1-{\bold E}^2+B^2}\right|}.
\eear
After taking ${\cal T}_2=1$, shapes of potential $U_{\rm EM}(T)$
for various values of electric field $E_2$ are depicted in Fig. 9.
The smaller electric field $E_2$ becomes, the more concave function in potential $U_{\rm EM}(T)$.
Then, profiles of tachyonic field $T(x)$ for various electric field $E_2$ and
energy density are depicted in Fig. 10 and Fig. 11.
Then, the configurational entropy $S_{\rm C,EM}$ is depicted in Fig. 12.
As electric field $E_2$ grows up,
$S_{\rm C,EM}$ starts at the minimum value ($S_{\rm C,EM}=3.8446$)
and saturates to the maximum value ($S_{\rm C,EM}=3.8963$).

\begin{figure}[!htbp]
\begin{center}
{\includegraphics[width=8cm]{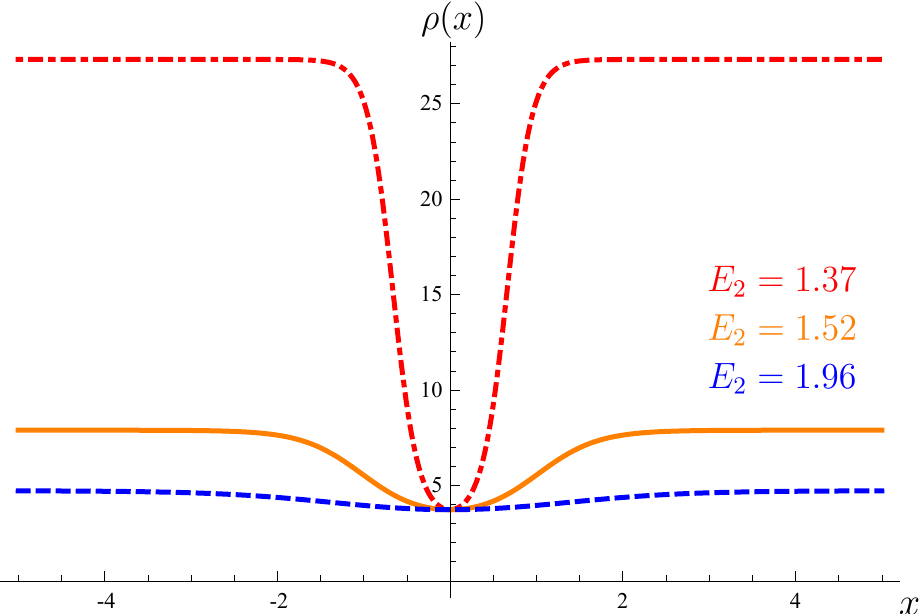}}
\end{center}
\vspace{-0.6cm}
\caption{{\footnotesize Plot of energy density $\rho(x)$ as
the function of the position $x$
(red dotted-dashed curve for
electric field $E_2=1.37$, orange solid curve for electric field $E_2
=1.52$, blue dashed curve for electric field $E_2=1.96$,
respectively).}}
\label{figXII}
\end{figure}

\begin{figure}[!htbp]
\begin{center}
{\includegraphics[width=8cm]{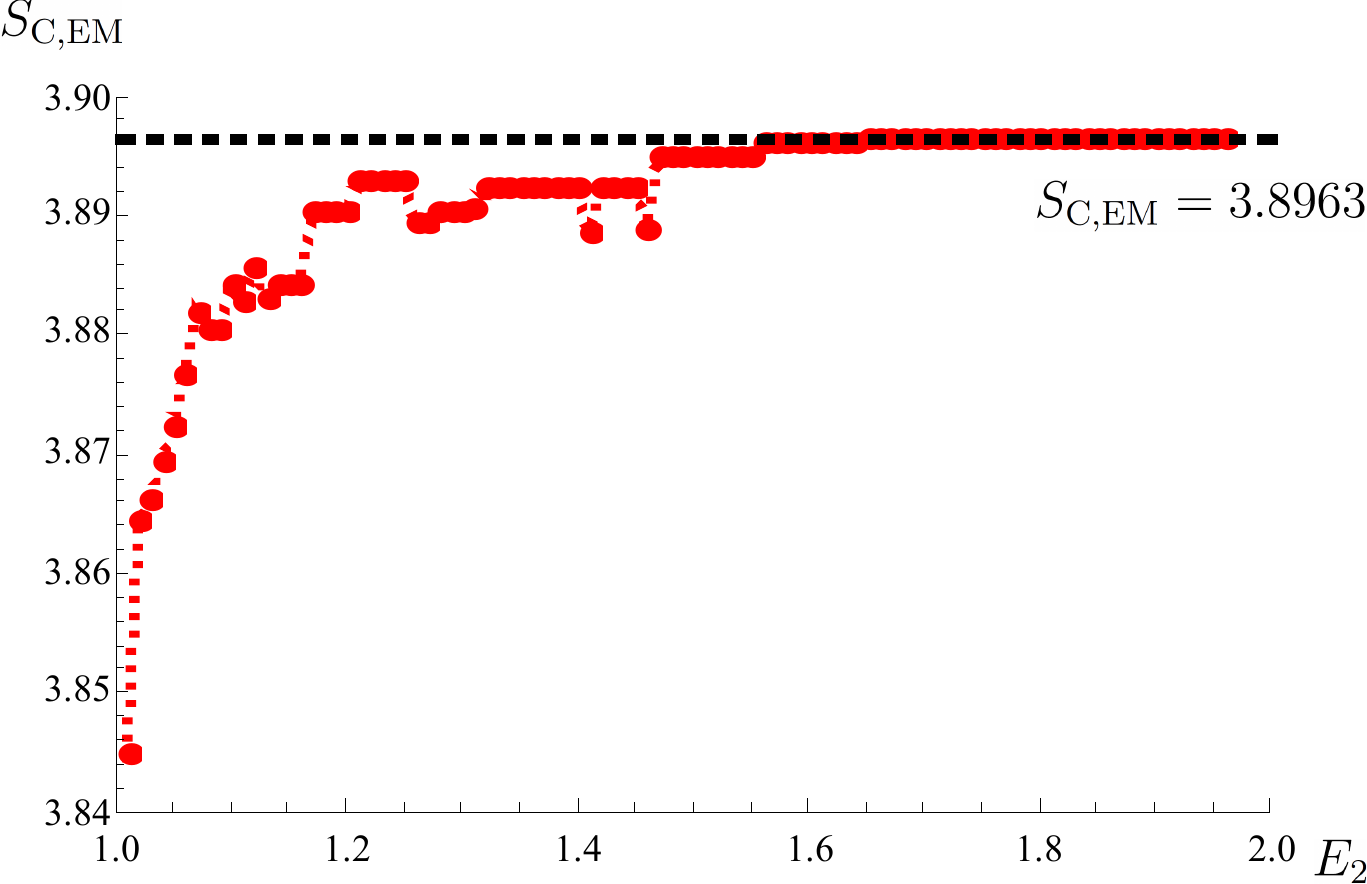}}
\end{center}
\vspace{-0.6cm}
\caption{{\footnotesize Plot of tachyon configurational entropy with electromagnetic field $S_{\rm C,EM}$
as the function of electric field $E_2$.}}
\label{figXIII}
\end{figure}

\begin{figure}[!htbp]
\begin{center}
{\includegraphics[width=8cm]{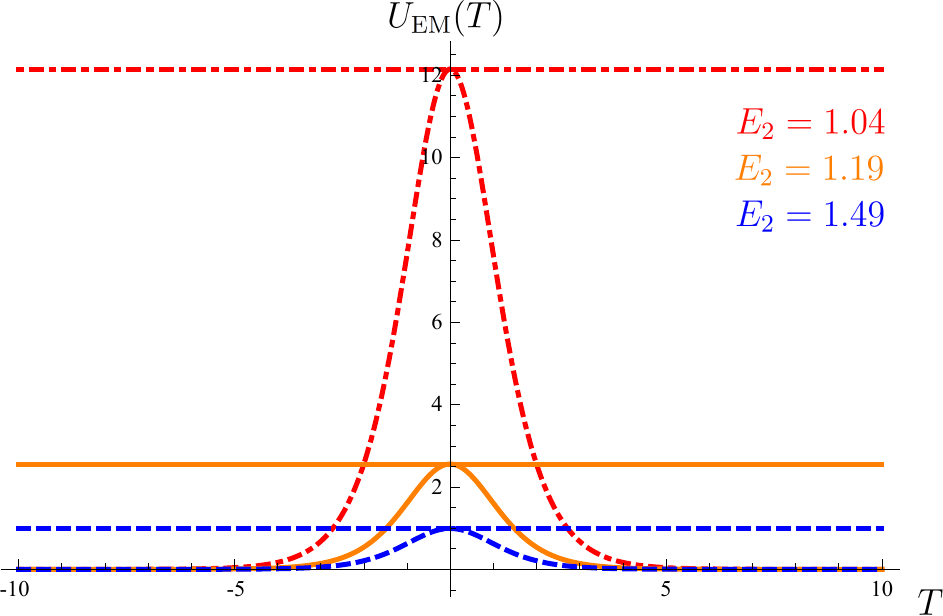}}
\end{center}
\vspace{-0.6cm}
\caption{{\footnotesize Plot of potential $U(T)$ as
the function of the tachyon field $T$ (red dotted-dashed curve for
electric field $E_2=1.04$, orange solid curve for electric
field $E_2 =1.19$, blue dashed curve for electric field
$E_2=1.49$, respectively).}}
\label{figXIV}
\end{figure}

(ii) When ${\cal E}_{\rm EM}=U_{\rm EM}(0)$ $(1-{\bold
E}^2+B^2=({\cal T}_2E_1/\Pi_1)^2$, as shown in Fig 13.), there are
the trivial ontop solution ($T(x)=0$) and nontrivial tachyon
half-kink solutions as shown in Fig 14. Then, the charge density is
given as
\bear
\rho&=&\frac{\Pi_1(B^2E_2^2-E_1^2)}{E_1(E_2^2-1)}\nonumber\\
&&-\frac{{\cal T}_2^2E_1}{\Pi_1(E_2^2-1)}\frac{1}{1+\exp(2x/\sigma)}.
\eear
After taking ${\cal T}_2=1$, shapes of potential $U_{\rm EM}(T)$
for various values of electric field $E_2$ are depicted in Fig. 13.
The smaller electric field $E_2$ becomes , the more concave function in potential $U_{\rm EM}(T)$.
Then, profiles of tachyon filed  $T(x)$ for various electric field $E_2$ and
energy density are depicted in Fig. 14 and Fig. 15.
Then, the configurational entropy $S_{\rm C,EM}$ is depicted in Fig. 16.
As electric field $E_2$ grows up,
$S_{\rm C,EM}$ starts at the minimum value ($S_{\rm C,EM}=0.0002618$)
and saturates to the maximum value ($S_{\rm C,EM}=3.7719$).

\begin{figure}[!htbp]
\begin{center}
{\includegraphics[width=8cm]{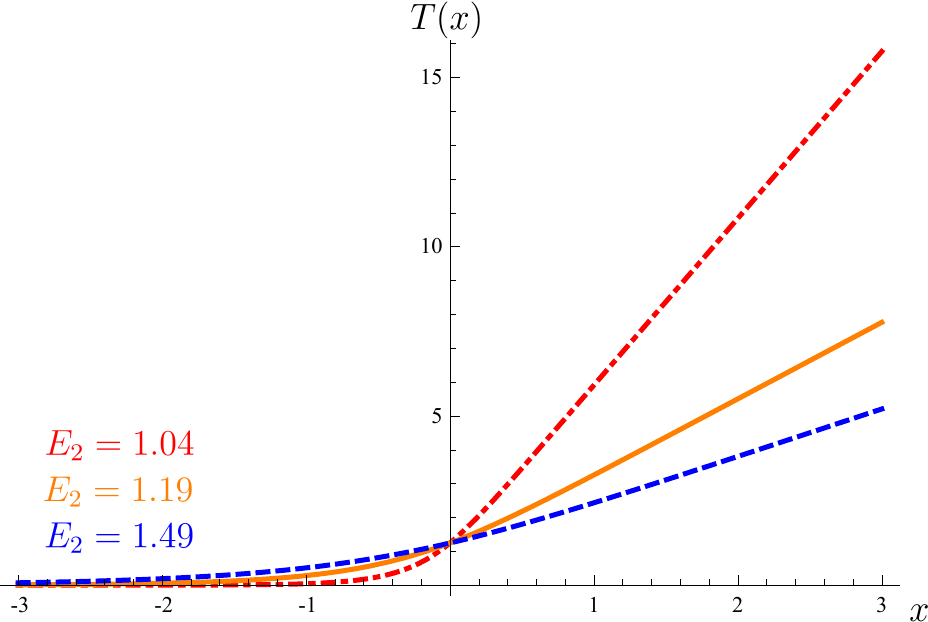}}
\end{center}
\vspace{-0.6cm}
\caption{{\footnotesize Plot of the tachyon field
$T(x)$ as the function of the position $x$ (red dotted-dashed curve
for electric field $E_2=1.04$, orange solid curve for
electric field $E_2 =1.19$, blue dashed curve for electric field
$E_2=1.49$, respectively).}}
\label{figXVI}
\end{figure}

\begin{figure}[!htbp]
\begin{center}
{\includegraphics[width=8cm]{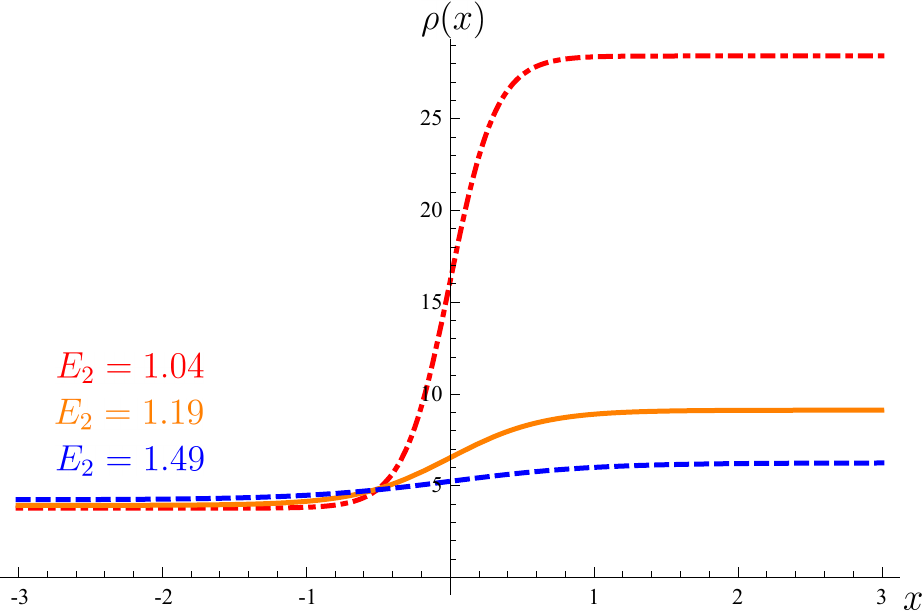}}
\end{center}
\vspace{-0.6cm}
\caption{{\footnotesize Plot of energy density
$\rho(x)$ as the function of the position $x$ (red dotted-dashed
curve for electric field $E_2=1.04$, orange solid curve for
electric field $E_2 =1.19$, blue dashed curve for electric field
$E_2=1.49$, respectively).}}
\label{figXVII}
\end{figure}

\begin{figure}[!htbp]
\begin{center}
{\includegraphics[width=8cm]{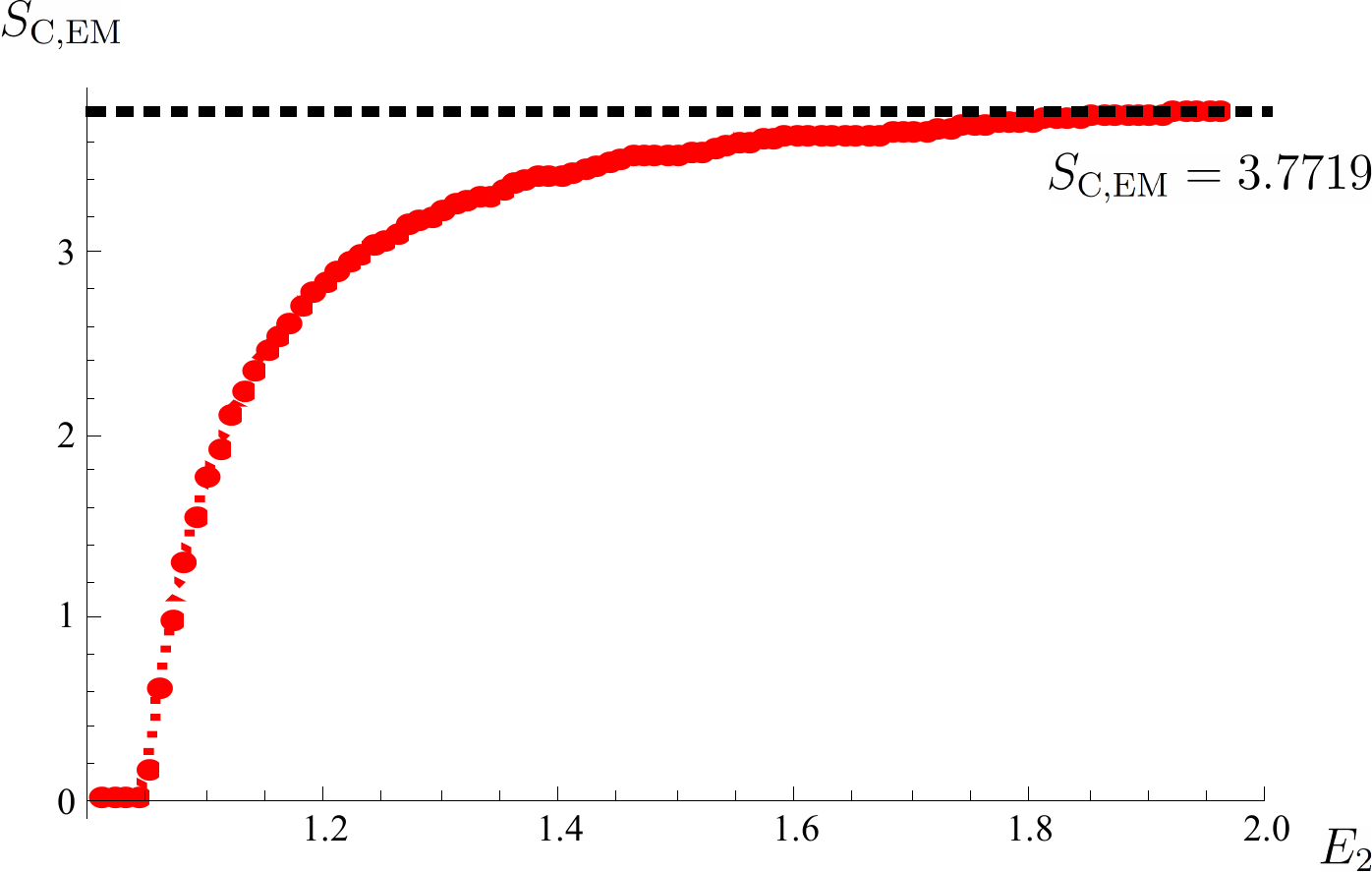}}
\end{center}
\vspace{-0.6cm}
\caption{{\footnotesize Plot of tachyon
configurational entropy with electromagnetic field $S_{\rm
C,EM}$ as the function of electric field $E_2$.}}
\label{figXVIII}
\end{figure}

(iii) When ${\cal E}_{\rm EM}>U_{\rm EM}(0)$
$(1-{\bold E}^2+B^2>({\cal T}_2E_1/\Pi_1)^2$, as shown in Fig 17.),
there is hybrid of two half-kink solutions joined at the origin,
as shown in Fig 18.
Then, the charge density is given as
\bear
\rho&=&\frac{\Pi_1(B^2E_2^2-E_1^2)}{E_1(E_2^2-1)}\nonumber\\
&&-\frac{{\cal T}_2^2E_1}{\Pi_1(E_2^2-1)}\frac{1}{1+(1-\chi^2)\sinh^2(x/\sigma)}.
\eear

\begin{figure}[!htbp]
\begin{center}
{\includegraphics[width=8cm]{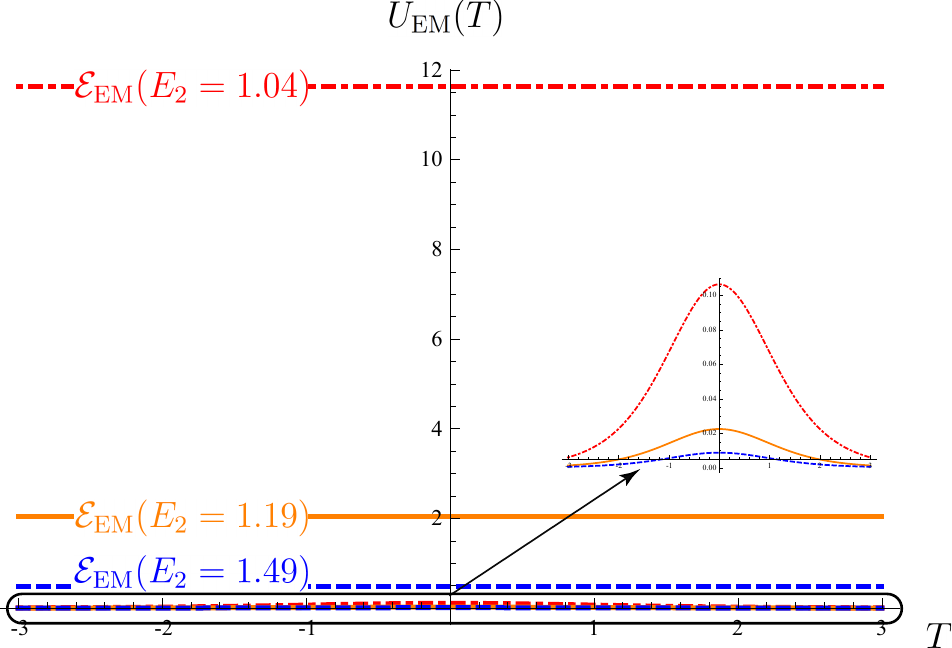}}
\end{center}
\vspace{-0.6cm}
\caption{{\footnotesize Plot of potential $U(T)$ as
the function of the tachyon field $T$ (red dotted-dashed curve for
electric field $E_2=1.04$, orange solid curve for electric
field $E_2 =1.19$, blue dashed curve for electric field
$E_2=1.49$, respectively).}}
\label{figXIX}
\end{figure}

\begin{figure}[!htbp]
\begin{center}
{\includegraphics[width=8cm]{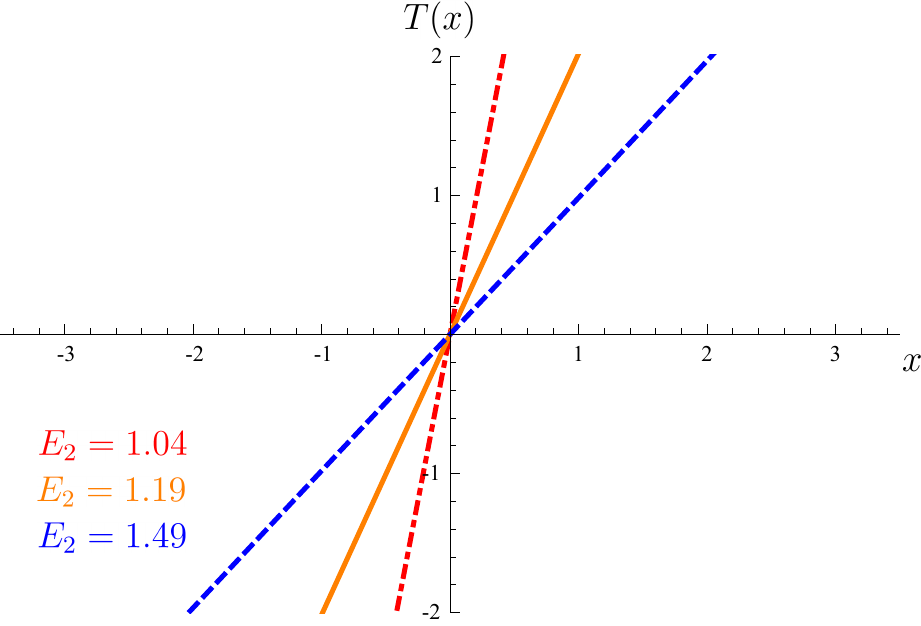}}
\end{center}
\vspace{-0.6cm}
\caption{{\footnotesize Plot of the tachyon field
$T(x)$ as the function of the position $x$ (red dotted-dashed curve
for electric field $E_2=1.04$, orange solid curve for
electric field $E_2 =1.19$, blue dashed curve for electric field
$E_2=1.49$, respectively).}}
\label{figXXI}
\end{figure}

\begin{figure}[!htbp]
\begin{center}
{\includegraphics[width=8cm]{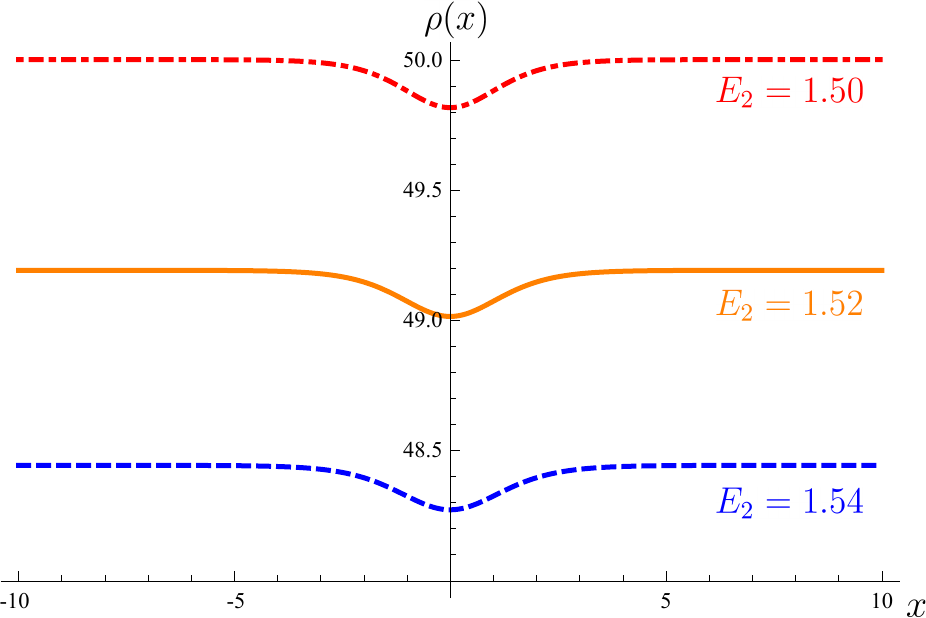}}
\end{center}
\vspace{-0.6cm}
\caption{{\footnotesize Plot of energy density
$\rho(x)$ as the function of the position $x$ (red dotted-dashed
curve for electric field $E_2=1.50$, orange solid curve for
electric field $E_2 =1.52$, blue dashed curve for electric field
$E_2=1.54$, respectively).}}
\label{figXXII}
\end{figure}

\begin{figure}[!htbp]
\begin{center}
{\includegraphics[width=8cm]{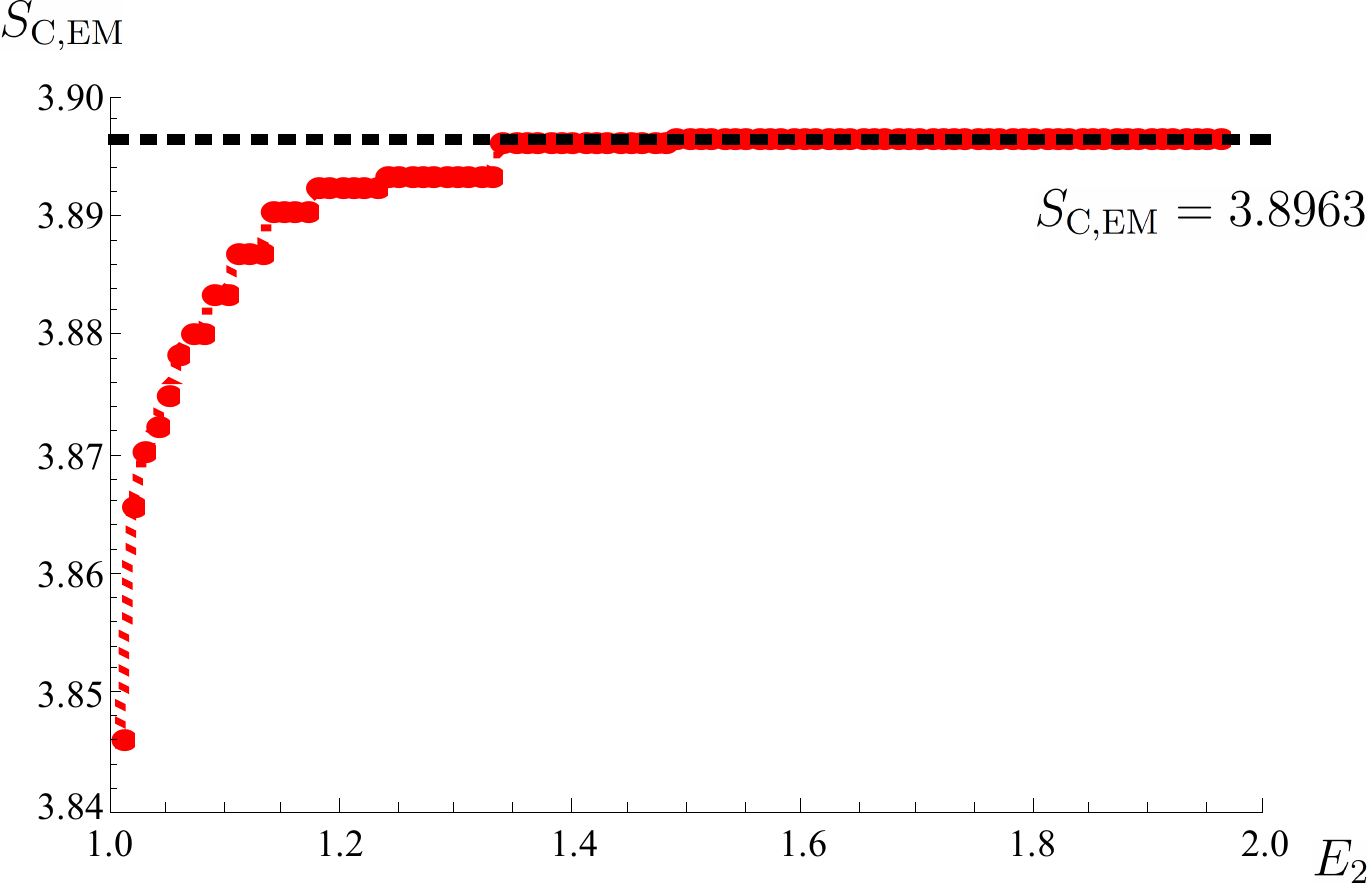}}
\end{center}
\vspace{-0.6cm}
\caption{{\footnotesize Plot of tachyon
configurational entropy with electromagnetic field $S_{\rm
C,EM}$ as the function of electric field $E_2$.}}
\label{figXXIII}
\end{figure}

After taking ${\cal T}_2=1$, shapes of potential $U_{\rm EM}(T)$ for
various values of electric field $E_2$ are depicted in Fig. 13.
The smaller electric field $E_2$ becomes, the more concave
function in potential $U_{\rm EM}(T)$. Then, profiles of tachyon
filed  $T(x)$ for various electric field $E_2$ and
energy density are depicted in Fig. 18. and Fig. 19. Then, the
configurational entropy $S_{\rm C,EM}$ is depicted in Fig. 20. As
electric field $E_2$ grows up, $S_{\rm C,EM}$ starts at the
minimum value ($S_{\rm C,EM}=3.8459$) and saturates to the maximum
value ($S_{\rm C,EM}=3.8963$).

\newpage
\section{Conclusion}
We considered the tachyonic system coupled to Born-Infeld
electromagnetism and investigated its configurational entropy. It
was found that the configurational entropy saturates to the maximum
value as the pressure of pure tachyonic field $-p_1$ grows up. We
also showed that the configurational entropy in the presence of
electromagnetic fields saturates to the maximum value as
electric field $E_2$ increases. Interestingly, when the magnetic field is turned off,
the magnitude of electric field $E$ reaches the critical value.
Then, the configurational entropy has the global minimum,
which is related to the predominant tachyonic states.

\section*{Acknowledgements}
This work was supported by Basic Science Research Program through
the National Research Foundation of Korea (NRF) funded by the
Ministry of Education, Science and Technology (NRF-2018R1D1A1B07049451).


\begin{thebibliography}{100}
\bibitem{Sen:1998sm}
  A.~Sen,
  JHEP {\bf 9808}, 012 (1998)
  [hep-th/9805170].

\bibitem{Sen:2002nu}
  A.~Sen,
  JHEP {\bf 0204}, 048 (2002)
  [hep-th/0203211].

\bibitem{Sen:2002in}
  A.~Sen,
  JHEP {\bf 0207}, 065 (2002)
  [hep-th/0203265].


\bibitem{Sen:2002an}
  A.~Sen,
  Mod.\ Phys.\ Lett.\ A {\bf 17}, 1797 (2002)
  [hep-th/0204143].

\bibitem{Cline:2002it}
  J.~M.~Cline, H.~Firouzjahi and P.~Martineau,
  JHEP {\bf 0211}, 041 (2002)
  [hep-th/0207156].

\bibitem{Felder:2002sv}
  G.~N.~Felder, L.~Kofman and A.~Starobinsky,
  JHEP {\bf 0209}, 026 (2002)
  [hep-th/0208019].

\bibitem{Felder:2004xu}
  G.~N.~Felder and L.~Kofman,
  Phys.\ Rev.\ D {\bf 70}, 046004 (2004)
  [hep-th/0403073].

\bibitem{Barnaby:2004nk}
  N.~Barnaby,
  JHEP {\bf 0407}, 025 (2004)
  [hep-th/0406120].

\bibitem{Lambert:2003zr}
  N.~D.~Lambert, H.~Liu and J.~M.~Maldacena,
  JHEP {\bf 0703}, 014 (2007)
  [hep-th/0303139].

\bibitem{Kluson:2003qk}
  J.~Kluson,
  JHEP {\bf 0401}, 019 (2004)
  [hep-th/0312086].

\bibitem{Kim:2003ina}
  C.~j.~Kim, Y.~b.~Kim and C.~O.~Lee,
  JHEP {\bf 0305}, 020 (2003)
  [hep-th/0304180]].

\bibitem{Kim:2003ma}
  C.~Kim, Y.~Kim, O.~K.~Kwon and C.~O.~Lee,
  JHEP {\bf 0311}, 034 (2003)
  [hep-th/0305092].

\bibitem{Sen:2003bc}
  A.~Sen,
  Phys.\ Rev.\ D {\bf 68}, 106003 (2003)
  [hep-th/0305011].

\bibitem{Gleiser:2011di}
  M.~Gleiser and N.~Stamatopoulos,
  Phys.\ Lett.\ B {\bf 713}, 304 (2012)
  [arXiv:1111.5597 [hep-th]].

\bibitem{Gleiser:2012tu}
  M.~Gleiser and N.~Stamatopoulos,
  Phys.\ Rev.\ D {\bf 86}, 045004 (2012)
  [arXiv:1205.3061 [hep-th]].


\bibitem{Bernardini:2016hvx}
  A.~E.~Bernardini and R.~da Rocha,
  Phys.\ Lett.\ B {\bf 762}, 107 (2016)
  [arXiv:1605.00294 [hep-th]].


\bibitem{Bernardini:2016qit}
  A.~E.~Bernardini, N.~R.~F.~Braga and R.~da Rocha,
  Phys.\ Lett.\ B {\bf 765}, 81 (2017)
  [arXiv:1609.01258 [hep-th]].


\bibitem{Braga:2017fsb}
  N.~R.~F.~Braga and R.~da Rocha,
  Phys.\ Lett.\ B {\bf 776}, 78 (2018)
  [arXiv:1710.07383 [hep-th]].


\bibitem{Barbosa-Cendejas:2018mng}
  N.~Barbosa-Cendejas, R.~Cartas-Fuentevilla, A.~Herrera-Aguilar, R.~R.~Mora-Luna and R.~da Rocha,
  Phys.\ Lett.\ B {\bf 782}, 607 (2018)
  [arXiv:1805.04485 [hep-th]].


\bibitem{Karapetyan:2018yhm}
  G.~Karapetyan,
  Phys.\ Lett.\ B {\bf 786}, 418 (2018)
  [arXiv:1807.04540 [nucl-th]].


\bibitem{Braga:2018fyc}
  N.~R.~F.~Braga, L.~F.~Ferreira and R.~Da Rocha,
  Phys.\ Lett.\ B {\bf 787}, 16 (2018)
  [arXiv:1808.10499 [hep-ph]].


\bibitem{Bernardini:2018uuy}
  A.~E.~Bernardini and R.~Da Rocha,
  Phys.\ Rev.\ D {\bf 98}, no. 12, 126011 (2018)
  [arXiv:1809.10055 [hep-th]].


\bibitem{Gleiser:2013mga}
  M.~Gleiser and D.~Sowinski,
  Phys.\ Lett.\ B {\bf 727}, 272 (2013)
  [arXiv:1307.0530 [hep-th]].

\bibitem{Gleiser:2014ipa}
  M.~Gleiser and N.~Graham,
  Phys.\ Rev.\ D {\bf 89}, no. 8, 083502 (2014)
  [arXiv:1401.6225 [astro-ph.CO]].

\bibitem{Gleiser:2015rwa}
  M.~Gleiser and N.~Jiang,
  Phys.\ Rev.\ D {\bf 92}, no. 4, 044046 (2015)
  [arXiv:1506.05722 [gr-qc]].

\bibitem{Casadio:2016aum}
  R.~Casadio and R.~da Rocha,
  Phys.\ Lett.\ B {\bf 763}, 434 (2016)
  [arXiv:1610.01572 [hep-th]].

\bibitem{Braga:2016wzx}
  N.~R.~F.~Braga and R.~da Rocha,
  Phys.\ Lett.\ B {\bf 767} (2017) 386
  [arXiv:1612.03289 [hep-th]].

\bibitem{Lee:2017ero}
  C.~O.~Lee,
  Phys.\ Lett.\ B {\bf 772}, 471 (2017)
  [arXiv:1705.09047 [gr-qc]].

\bibitem{Gleiser:2018kbq}
  M.~Gleiser, M.~Stephens and D.~Sowinski,
  Phys.\ Rev.\ D {\bf 97}, no. 9, 096007 (2018)
  [arXiv:1803.08550 [hep-th]].

\bibitem{German:2012rv}
  G.~German, A.~Herrera-Aguilar, D.~Malagon-Morejon, R.~R.~Mora-Luna and R.~da Rocha,
  JCAP {\bf 1302}, 035 (2013)
  [arXiv:1210.0721 [hep-th]].


















\end{thebibliography}
\end{document}